# IDBE - An Intelligent Dictionary Based Encoding Algorithm for Text Data Compression for High Speed Data Transmission Over Internet

V.K. Govindan[1]    B.S. Shajee mohan[2]

**Abstract**
Compression algorithms reduce the redundancy in data representation to decrease the storage required for that data. Data compression offers an attractive approach to reducing communication costs by using available bandwidth effectively.  Over the last decade there has been an unprecedented explosion in the amount of digital data transmitted via the Internet, representing text, images, video, sound, computer programs, etc.  With this trend expected to continue, it makes sense to pursue research on developing algorithms that can most effectively use available network bandwidth by maximally compressing data. This research paper is focused on addressing this problem of lossless compression of text files.  Lossless compression researchers have developed highly sophisticated approaches, such as Huffman encoding, arithmetic encoding, the Lempel-Ziv family, Dynamic Markov Compression (DMC), Prediction by Partial Matching (PPM), and Burrows-Wheeler Transform (BWT) based algorithms.  However, none of these methods has been able to reach the theoretical best-case compression ratio consistently, which suggests that better algorithms may be possible.  One approach for trying to attain better compression ratios is to develop new compression algorithms.  An alternative approach, however, is to develop intelligent, reversible transformations that can be applied to a source text that improve an existing, or backend, algorithm's ability to compress.  The latter strategy has been explored here.
Michael Burrows and David Wheeler recently released the details of a transformation function that opens the door to some revolutionary new data compression techniques. The Burrows-Wheeler Transform, or *BWT*, transforms a block of data into a format that is extremely well suited for compression. The block sorting algorithm they developed works by applying a reversible transformation to a block of input text. The transformation does not itself compress the data, but reorders it to make it easy to compress with simple algorithms such as move to front encoding.
The basic philosophy adopted by us in this paper is  to preprocess the text and transform it into some intermediate form which can be compressed with better efficiency and which exploits the natural redundancy of the language in making the transformation. A strategy called Intelligent Dictionary Based Encoding (IDBE) is discussed to achieve this. It has been observed that a preprocessing of the text prior to  conventional  compression   will  improve  the  compression  efficiency  much  better. The intelligent dictionary based encryption provides the required security.

1.Department of Computer Science and Engineering, National Institute of Technology, Calicut, Kerala, India, vkg@nitc.ac.in 2. Department of Computer Science and Engineering, L.B.S. College of Engineering, Kasaragode, Kerala, India, shajeemohan@yahoo.com

## 1.0. Related Work and Background

In the last decade, we have seen an unprecedented explosion of textual information through the use of the Internet, digital library and information retrieval system. It is estimated that by the year 2004 the National Service Provider backbone will have an estimated traffic around 30000Gbps and that the growth will continue to be 100% every year. The text data competes for 45% of the total Internet traffic. A number of sophisticated algorithms have been proposed for lossless text compression of which BWT and PPM out perform the classical algorithms like Huffman, arithmetic and LZ families of Gzip and Unix compress. The BWT is an algorithm that takes a block of data and rearranges it using a sorting algorithm. The resulting output block contains exactly the same data elements that it started with, differing only in their ordering. The transformation is reversible, meaning the original ordering of the data elements can be restored with no loss of fidelity.

The BWT is performed on an entire block of data at once. Most of today's familiar lossless compression algorithms operate in streaming mode, reading a single byte or a few bytes at a time. But with this new transform, we want to operate on the largest chunks of data possible. Since the BWT operates on data in memory, you may encounter files too big to process in one fell swoop. In these cases, the file must be split up and processed a block at a time. The output of the BWT transform is usually piped through a move-to-front stage, then a run length encoder stage, and finally an entropy encoder, normally arithmetic or Huffman coding. The actual command line to perform this sequence will look like this:

```
BWT < input-file | MTF | RLE | ARI > output-file
```

The decompression is just the reverse process and look like this

```
UNARI input-file | UNRLE | UNMTF | UNBWT  > output-file
```

An alternate approach to this is to perform a lossless, reversible transformation to a source file prior to applying an existing compression algorithm. The transformation is designed to make it easier to compress the source file. The star encoding is generally used for this type of pre processing transformation of the source text. Star-encoding works by creating a large dictionary of commonly used words expected in the input files. The dictionary must be prepared in advance, and must be known to the compressor and decompressor.

Each word in the dictionary has a star-encoded equivalent, in which as many letters a possible are replaced by the '*' character. For example, a commonly used word such *the* might be replaced by the string *t\*\**. The star-encoding transform simply replaces every occurrence of the word the in the input file with *t\*\**.

Ideally, the most common words will have the highest percentage of '*' characters in their encodings. If done properly, this means that transformed file will have a huge number of '*' characters. This ought to make the transformed file more compressible than the original plain text. The existing star encoding does not provide any compression as such but provide the input text a better compressible format for a later stage compressor. The star encoding is very much weak and vulnerable to attacks.

As an example, a section of text from Project Guttenburg's version of *Romeo and Juliet* looks like this in the original text:

```
But soft, what light through yonder window breaks?
It is the East, and Iuliet is the Sunne,
Arise faire Sun and kill the enuious Moone,
Who is already sicke and pale with griefe,
That thou her Maid art far more faire then she
```

Running this text through the star-encoder yields the following text:

```
B** *of*, **a* **g** *****g* ***d*r ***do* b*e***?
It *s *** E**t, **d ***i** *s *** *u**e,
A***e **i** *un **d k*** *** e****** M****,
*ho *s a****** **c*e **d **le ***h ****fe,
***t ***u *e* *ai* *r* f*r **r* **i** ***n s**
```

We can clearly see that the encoded data has exactly the same number of characters, but is dominated by stars. A later modification for star encoding called LIPT (Length Index Preserving Transform) whose genesis can be traced to several other similar transforms developed by the M-5 Research group at the Department of Computer Science, University of Central Florida, should also be mentioned at this juncture

## 2.0 An Intelligent Dictionary Based Encoding.

In these circumstances we propose a better encoding strategy, which will offer higher compression ratio. The objective of this paper is to develop a better transformation yielding greater compression. The basic philosophy of compression is to transform text in to some intermediate form, which can be compressed with better efficiency, by exploiting the natural redundancy of the language in making this transformation. We have explained the basic approach of our compression method in the previous sentence but let us use the same sentence as an example to explain the point further. ***The objectiv of tis paper is to develop a bettr transfomation yielding greater compresion. The basic philosopy of comprssion is to transfom text in to some intermedate form,..*** Most people will have no problem to read it. This is because our visual perception system recognizes each word with an approximate signature pattern for the word opposed to an actual and exact sequence of letters and we have a dictionary in our brain, which associates each misspelled word with a corresponding, correct word. The signatures for the word for computing machinery could be arbitrary as long as they are unique. The algorithm we developed is a two step process consisting

Step1: Make an intelligent dictionary

Step2: Encode the input text data

The entire process can be summerised as follows.

### 2.1 Encoding Algorithm

Start encode with argument input file ***inp***

    A. Read the dictionary and store all words and their codes in a table

B. While *inp* is not empty

   1. Read the characters from *inp* and form tokens.

   2. If the token is longer than 1 character, then

1. Search for the token in the table
2. If it is not found,
    1. Write the token as such in to the output file.

   Else

   1. Find the length of the code for the word.
   2. The actual code consists of the length concatenated with the code in the table, the length serves as a marker while decoding and is represented by the ASCII characters 251 to254 with 251 representing a code of length 1, 252 length 2 and so on.
3. Write the actual code into the output file.
4. read the next character and neglect the it if it is a space. If it is any other character, make it the first character of the next token and go back to **B,** after inserting a marker character (ASCII 255) to indicate the absence of a space.

   Endif

   Else

1. Write the 1 character token
2. If the character is one of the ASCII characters 251 –255, write the character once more so as to show that it is part of the text and not a marker

Endif

End (While)

  C. Stop.

**2.2. Dictionary Making Algorithm**

Start MakeDict with multiple source files as input

1. Extract all words from input files.
2. If a word is already in the table increment the number of occurrence by 1, otherwise add it to the table and set the number occurrence to 1.
3. Sort the table by frequency of occurrences in descending order.
4. Start giving codes using the following method:

   i). Give the first 218 words the ASCII characters 33 to 250 as the code.

   ii). Now give the remaining words each one permutation of two of the ASCII characters (in the range 33 – 250), taken in order. If there are any remaining words give them each one permutation of three of the ASCII characters and finally if required permutation of four characters.

5. Create a new table having only words and their codes. Store this table as the Dictionary in a file.

6. Stop.

As an example, to demonstrate this a section of the text from Canterbury corpus version of *bible.txt* which looks like this in the original text:

> In the beginning God created the heaven and the earth. And the earth was without form, and void; and darkness was upon the face of the deep. And the Spirit of God moved upon the face of the waters.
>
> And God said, Let there be light: and there was light.
>
> And God saw the light, that it was good: and God divided the light from the darkness.
>
> And God called the light Day, and the darkness he called Night. And the evening and the morning were the first day.
>
> And God said, Let there be a firmament in the midst of the waters, and let it divide the waters from the waters.
>
> And God made the firmament, and divided the waters which were under the firmament from the waters which were above the firmament: and it was so.
>
> And God called the firmament Heaven. And the evening and the morning were the second day.

Running the above text through our Intelligent Dictionary Based Encoder (IDBE), which we have implemented in C ++, yields the following text:

û©û!ü%;ûNü'Œû!ü"ƒû"û!û˜ÿ. û*û!û˜û5ü"8ü"}ÿ, û"ü2Óÿ; û"ü%Lû5ûYû!ü"nû#û!ü&"ÿ. û*û!ü%Ìû#ûNü&ÇûYû!ü"nû#û!ü#Éÿ.

û*ûNûAÿ, ü"¿û]û.ü"'ÿ: û"û]û5ü"'ÿ.

û*ûNü"Qû!ü"'ÿ, û'û1û5û²ÿ: û"ûNü(Rû!ü"'û;û!ü%Lÿ.

û*ûNûóû!ü"ü%…ÿ, û"û!ü%Lû-ûóü9[ÿ. û*û!ü'·û"û!ü#'ûSû!ûºûvÿ.

û*ûNûAÿ, ü"¿û]û.û&ü6  û%û!ü#?û#û!ü#Éÿ, û"û«û1ü,-û!ü#Éû;û!ü#Éÿ.

û*ûNû,û!ü6  ÿ, û"ü(Rû!ü#Éû:ûSü"2û!ü6  û;û!ü#Éû:ûSü",û!ü6  ÿ: û"û1û5ûeÿ.

û*ûNûóû!ü6  ü#Wÿ. û*û!ü'·û"û!ü#'ûSû!ü"ßûvÿ.

It is clear from the above sample data that the encoded text provide a better compressive format for a conventional BWT based compression module. The research finding and comparisons are given in tabular format in the next section.

## 3. 0. Performance analysis

The performance issues such as Bits Per Character (BPC) and conversion time are compared for the three cases i.e., simple BWT, BWT with Star encoding and BWT with our proposed Intelligent Dictionary Based Encoding (IDBE). The results are shown graphically and prove that BWT with IDBE out performs all other techniques in compression ratio and speed of compression (conversion time)..

**Table 1.0** BPC comparison of simple BWT, BWT with *Encode and BWT with IDBE in Calgary corpuses

| Calgary corpuses | | | | | | | |
|---|---|---|---|---|---|---|---|
| File Names | File size Kb | BWT | | BWT with *Encode | | BWT with IDBE | |
| | | BPC | Time (Secs) | BPC | Time (Secs) | BPC | Time (Secs) |
| bib | 108.7 | 2.11 | 1 | 1.93 | 6 | 1.69 | 4 |
| book1 | 750.8 | 2.85 | 11 | 2.74 | 18 | 2.36 | 11 |
| book2 | 596.5 | 2.43 | 9 | 2.33 | 14 | 2.02 | 10 |
| geo | 100.0 | 4.84 | 2 | 4.84 | 6 | 5.18 | 5 |
| news | 368.3 | 2.83 | 6 | 2.65 | 10 | 2.37 | 7 |
| paper1 | 51.9 | 2.65 | 1 | 1.59 | 5 | 2.26 | 3 |
| paper2 | 80.3 | 2.61 | 2 | 2.45 | 5 | 2.14 | 4 |
| paper3 | 45.4 | 2.91 | 2 | 2.60 | 6 | 2.27 | 3 |
| Paper4 | 13.0 | 3.32 | 2 | 2.79 | 5 | 2.52 | 3 |
| Paper5 | 11.7 | 3.41 | 1 | 3.00 | 4 | 2.8 | 2 |
| Paper6 | 37.2 | 2.73 | 1 | 2.54 | 5 | 2.38 | 3 |
| progc | 38.7 | 2.67 | 2 | 2.54 | 5 | 2.44 | 3 |
| progl | 70.0 | 1.88 | 1 | 1.78 | 5 | 1.70 | 3 |
| trans | 91.5 | 1.63 | 2 | 1.53 | 5 | 1.46 | 4 |

**Table 2.0** BPC comparison of simple BWT, BWT with *Encode and BWT with IDBE in Canterbury corpuses

| Cantebury corpuses | | | | | | | |
|---|---|---|---|---|---|---|---|
| File Names | File size Kb | BWT | | BWT with *Encode | | BWT with IDBE | |
| | | BPC | Time (Secs) | BPC | Time (Secs) | BPC | Time (Secs) |
| alice29.txt | 148.5 | 2.45 | 3 | 2.39 | 6 | 2.11 | 4 |
| Asyoulik.txt | 122.2 | 2.72 | 2 | 2.61 | 7 | 2.32 | 4 |
| cp.html | 24.0 | 2.6 | 1 | 2.27 | 4 | 2.13 | 3 |
| fields.c | 10.9 | 2.35 | 0 | 2.20 | 4 | 2.06 | 3 |
| grammar.lsp | 3.60 | 2.88 | 0 | 2.67 | 4 | 2.44 | 3 |
| kennedy.xls | 1005.6 | 0.810 | 10 | 0.823 | 17 | 0.976 | 17 |
| Icet10.txt | 416.8 | 2.38 | 7 | 2.25 | 12 | 1.87 | 7 |
| plrabn12.txt | 470.6 | 2.80 | 10 | 2.69 | 13 | 2.30 | 8 |
| ptt5 | 501.2 | 0.846 | 27 | 0.847 | 33 | 0.856 | 31 |
| sum | 37.3 | 2.80 | 2 | 2.75 | 4 | 2.89 | 4 |
| xrgs.1 | 4.1 | 3.51 | 1 | 3.32 | 4 | 2.93 | 2 |

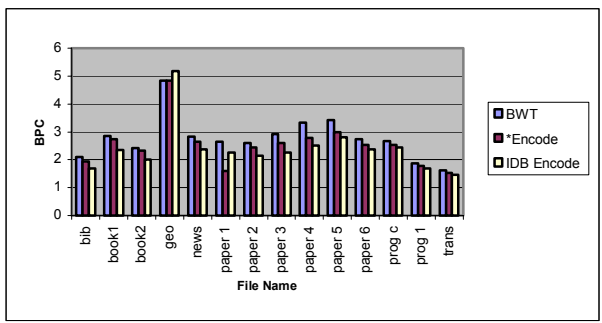 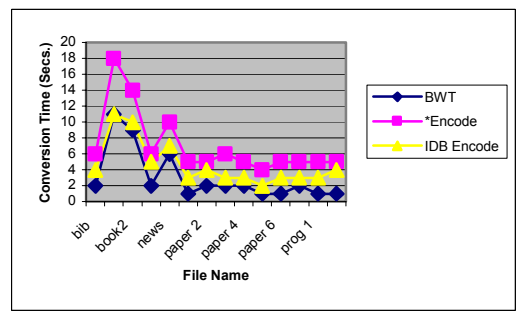

**Fig.1.0** BPC & Conversion time comparison of transform with BWT, BWT with *Encoding and BWT with IDBE for Calgary corpus files.

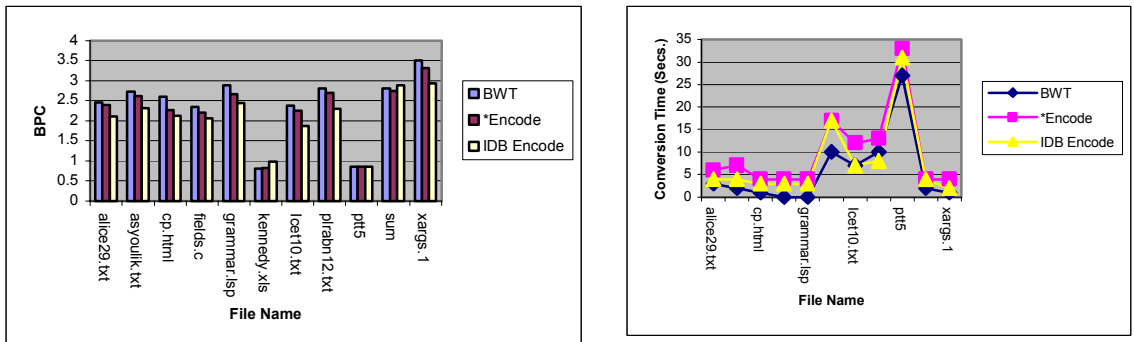

**Fig.2.0** BPC & Conversion time comparison of transform with BWT, BWT with *Encoding and BWT with IDBE for Canterbury corpus files.

## 4.0. Conclusion

In an ideal channel, the reduction of transmission time is directly proportional to the amount of compression. But in a typical Internet scenario with fluctuating bandwidth, congestion and protocols of packet switching, this does not hold true. Our results have shown excellent improvement in text data compression and added levels of security over the existing methods. These improvements come with additional processing required on the server/nodes.